\documentclass[12pt]{iopart}
\usepackage{here,epsfig,graphicx,graphics}
\usepackage[figuresright]{rotating}

\begin{document}

\title[Summary: Photons, Leptons and Heavy Quarks]{Summary of Experimental Results: Photons, Leptons and Heavy Quarks}

\author{Richard K. Seto}

\address{Physics Department 
University of California, Riverside, Riverside, CA 92521, USA}
\ead{richard.seto@ucr.edu}
\begin{abstract}
This is a summary of experimental results on photons, leptons, and
heavy quarks presented at Quark Matter 2008. A first measurement of
the bottom to charm contribution to the lepton spectrum has given
experimental indication for the suppression of charm and bottom.
Excess dileptons have been observed and studied by both NA60 and
PHENIX, which may arise from the early production of thermal dileptons
and/or the modification of mesons.

\end{abstract}


\section{Introduction}
The evidence for the production of a strongly interacting Quark Gluon
Plasma (sQGP) at RHIC is very strong, however coming into this conference
there are questions one would like to have answered. The hallmark
signatures for the sQGP have been the suppression of high momentum
particles, and the strong elliptic flow of hadrons. Surprisingly, heavy
quarks seen via their leptonic decays have shown a similar
behavior. Since both charm and bottom quarks decay leptonically, the
first question one would like to answer is whether this behavior is
limited to the charm quark, or includes the much heavier bottom quark
as well. Secondly, since strong elliptic flow indicates that the
system is thermalized, one is led to ask whether the ``black body''
radiation from this thermalized system in the form of thermal
dileptons can be seen. Dileptons can pass through the system
basically unhindered, hence they can give us information about the
time evolution of the system - both from the aforementioned thermal
radiation and from the decays of hadrons, or quasi-hadrons inside the
high temperature region which may allow us to observe hadron masses
and widths in a chirally symmetric phase. Finally we will ask whether
new results in the ongoing story of the J/$\psi$ can give us any
indication of deconfinement.

\section{Heavy Quarks}
\begin{figure}[hbt]
  \begin{center}
    \hspace*{-0.12in}
    \includegraphics[width=1.\linewidth]{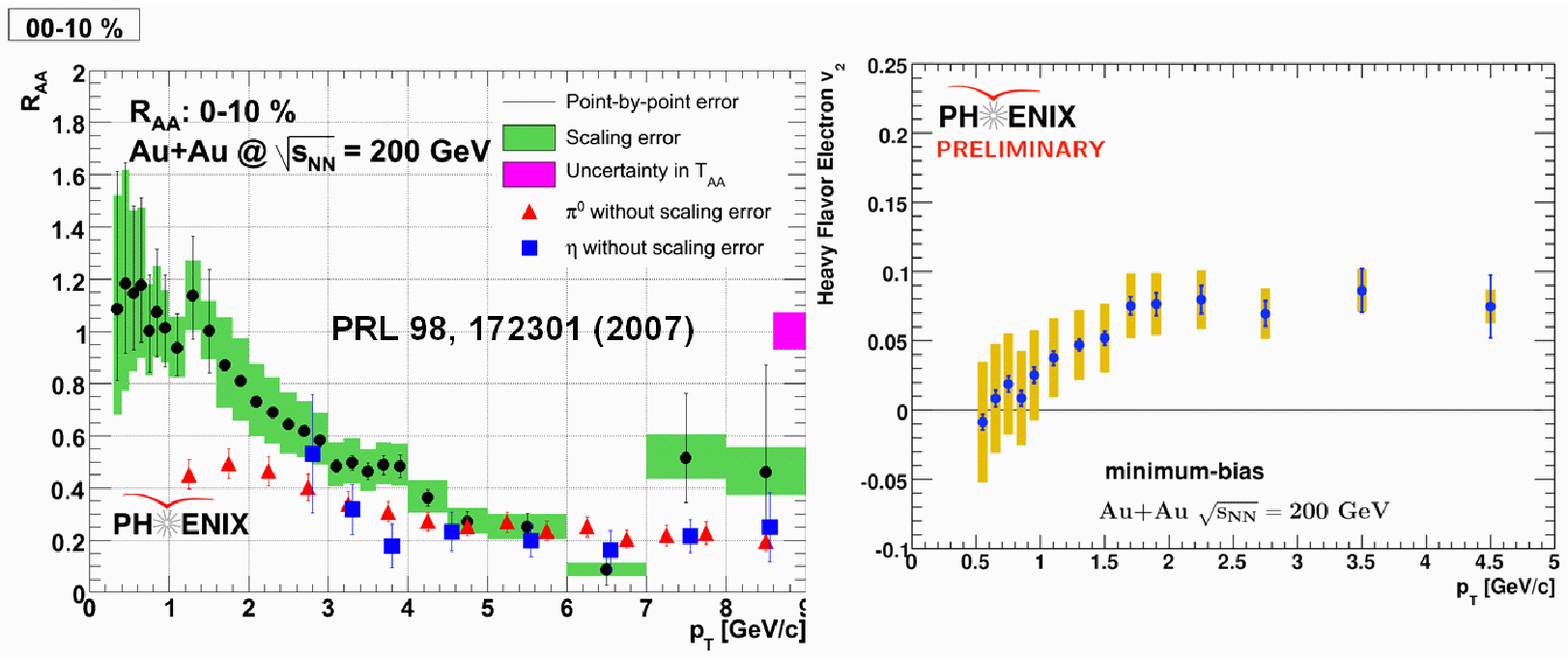}
  \end{center}\vspace*{-0.12in}
  \caption{\label{fig:cbsuppflow} 
     Left: R$_{AA}$ for non-photonic electrons and $\pi^0$s in central
Au+Au collisions. Right: v$_2$ for non photonic electrons in minimum bias Au+Au collisions. 
    }
\end{figure}

One of the most important early experimental results was the
suppression of high momentum particles. The standard explanation for
this effect is that high momentum quarks/gluons loose energy before
exiting the medium.  This signature is an indication of a very high
energy density; estimates range from 25 to 100 times the density of
nuclear matter. The strength of the elliptic flow leads one to believe
that thermalization occurs very early, probably
$\sim$1~fm/c\cite{Heinz:2004pj} though some recent estimates have been
much earlier than this. This leads to a picture of quarks, even quarks
of relatively high momentum, being thermalized rather quickly (or
nearly thermalized) because of the strength of the interaction they
experience. This could come from a large coupling constant, or from a
smaller coupling constant together with a turbulent
system\cite{Strickland:2007fm}. A striking, rather recent result, is
that the interactions are so strong that heavy quarks such as charm
also begin to thermalize and exhibit both suppression at high
transverse momentum, and elliptic flow.  Charm and bottom quarks, were
not expected to thermalize and flow with the medium because of their
mass - m$_c$ $\sim$ 1.3 GeV, m$_b$ $\sim$ 5 GeV. In addition, if the
suppression of high p$_T$ particles is a consequence of radiative
energy loss, then the ``dead-cone effect\cite{Dokshitzer:2001zm}''
would mean that heavy quarks should show less suppression.  Heavy
quarks have been observed using non-photonic electrons (electrons
spectra after the contribution of electrons from photon conversion and
Dalitz decays are subtracted) which come from the semi-leptonic decay
of both charm and bottom mesons. These indicate that heavy quarks are
suppressed at high p$_T$\cite{Adare:2006nq} (see
Fig.~\ref{fig:cbsuppflow}, left). An update of the non-photonic
electron flow is shown in Fig.~\ref{fig:cbsuppflow} (right) where the
v$_2$, now shown out to a p$_T$ of 5 GeV, appears to saturate at about
0.07.

\begin{figure}[hbt]
  \begin{center}
    \hspace*{-0.12in}
    \includegraphics[width=0.5\linewidth]{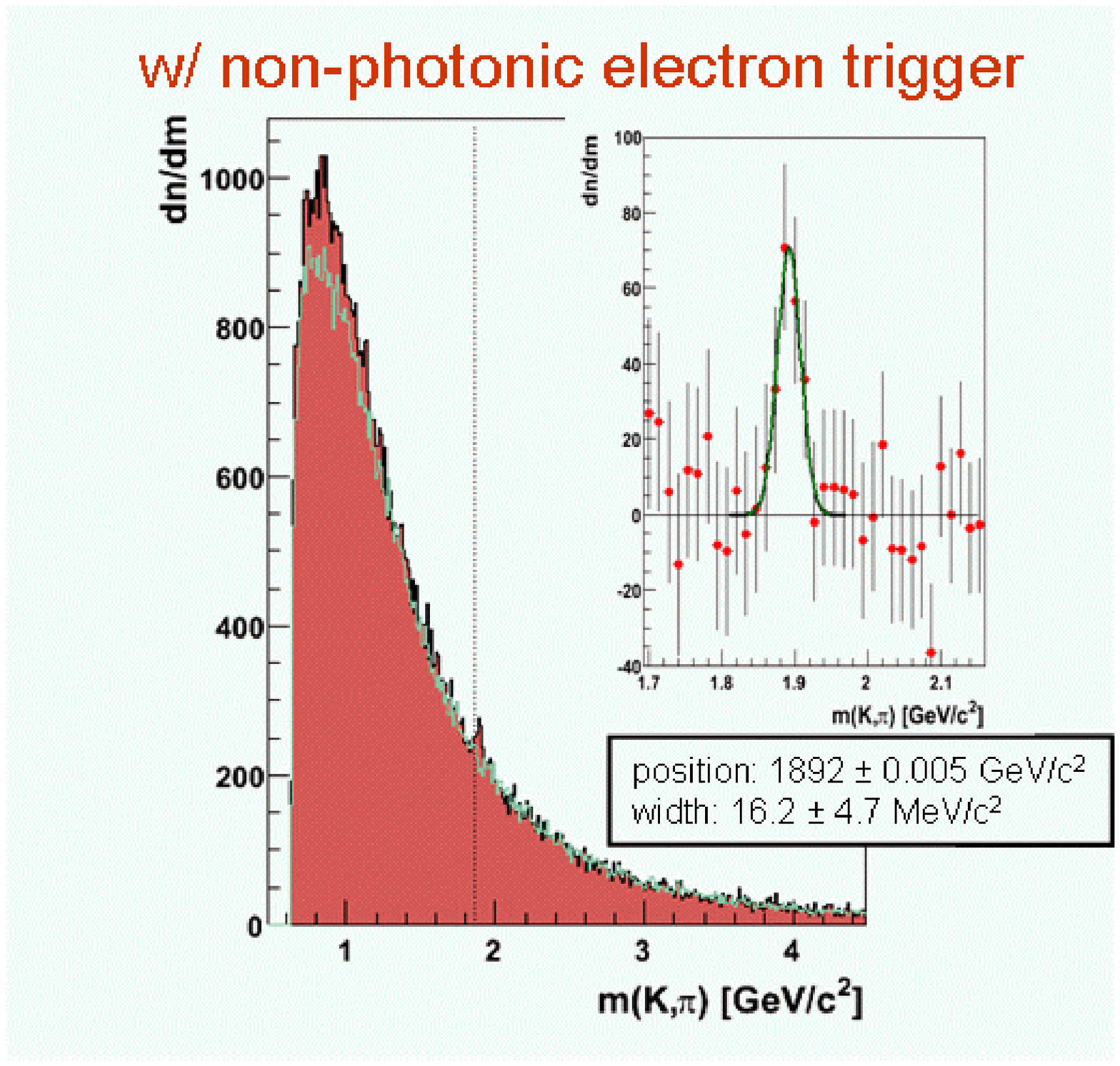}
    \includegraphics[width=0.5\linewidth]{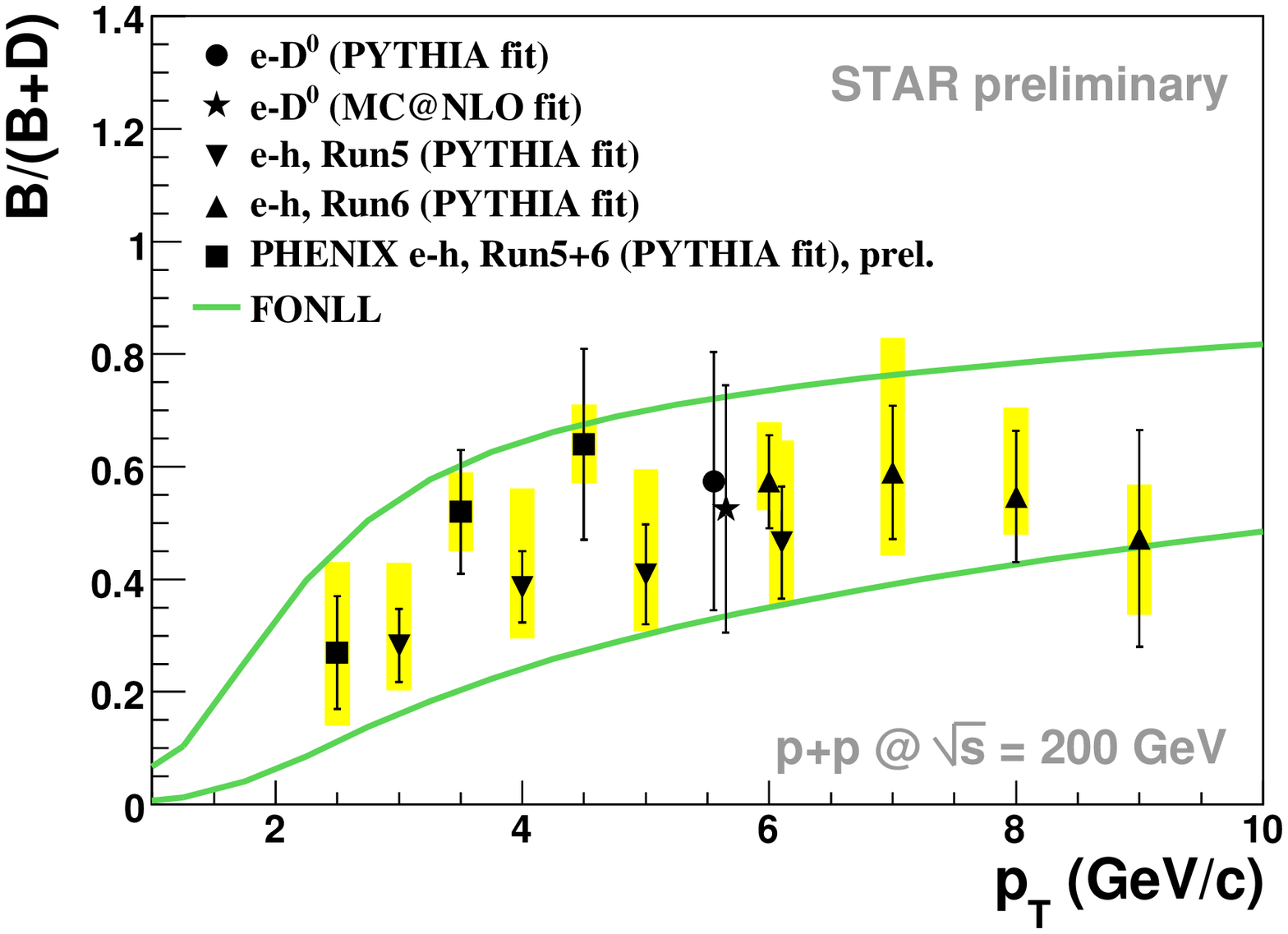}
  \end{center}\vspace*{-0.12in}
  \caption{\label{fig:cb} Left: The K$\pi$ invariant mass in 200 GeV
    p+p collisions using a non-photonic electron trigger as measured
    by STAR. The inset shows a background subtracted
    distribution. Right: B/(B+D) ratio as measured by STAR and PHENIX
    compared to FONLL calculations.  }
\end{figure}

Clearly charm flows and is suppressed at high p$_T$ - what about
bottom? Since heavy quarks are produced in the earliest moments of the
collision via perturbative processes, their initial yield and p$_T$
distribution in heavy ion collisions will be the same as in pp
collisions scaled by the number of binary collisions.  Both STAR and
PHENIX have now made a measurement of the b/c ratio in p+p collisions.
PHENIX uses a technique of examining the K+e invariant mass
distribution from the D$\rightarrow Ke\nu$ decay, subtracting off like
sign combinations and fitting the result to a combination of b and c
contributions.  STAR uses a similar method but in addition they
directly reconstruct the $D\rightarrow K\pi$ decay with an electron
tag (see Fig.~\ref{fig:cb}, left). Fig.~\ref{fig:cb} (right) shows the
D/(D+B) ratio obtained from the two experiments; the results agree
nicely with the FONLL expectations. The implication is that bottom is
responsible for a substantial part of the non-photonic electron signal
- about 50\% at p$_T$ greater than 3 or 4 GeV. Taken together with the
R$_{AA}$ and v$_2$ results in Au+Au, this indicates that the b quark
is at least partially thermalized in central Au+Au collisions and
accounts for a non-negligible amount of the suppression and elliptic
flow of non-photonic electrons. This is a very unexpected result since
the relaxation time of the b-quark is much longer than
that of the light quarks, because of its heavy mass.  This would
further imply that the strength of the interactions is strong - strong
enough to force the bottom quarks to move with the bulk of the sQGP,
hence non-perturbative calculations become necessary.  A Langevin
based model which includes resonant elastic scattering obtains a good
fit to both R$_{AA}$ and v$_2$ with a diffusion coefficient
D$_{HQ}$=(4 to 6)/2$\pi$T\cite{vanHees:2004gq}.

One can then make an estimate for $\eta$/s using transport models via
the diffusion coefficient for heavy quarks.  Moore and Teaney
calculate, using perturbative methods that D$_{HQ}$$\sim$6$\eta$/($\epsilon$+P)\cite{Moore:2004tg} and give a plausibility argument that the
result is true non-perturbatively.
Since at
$\mu_B$=0, $\epsilon$+P=Ts, one arrives at an answer for $\eta$/s
$\sim$ (1.3 to 2.0)/4$\pi$\cite{Adare:2006nq}, close the the
conjectured lower bound from AdS/CFT
calculations\cite{Policastro:2001yc} which assume extremely strong
interactions.

\section{Dileptons and Direct Photons}
As we explore the properties of the sQGP, we wish to obtain
information from early times in the evolution of the collision.  Since
they do not experience the strong interaction, leptons and photons are
ideal messengers of information from deep within the fireball.  They
can give us vital information on two fronts. The first, via leptonic
decays of hadrons (or at least quasi-hadrons), is
information on the characteristics of hadrons in the high temperature
environment. They will tell us of modifications to the vacuum
hadronic states - in particular - whether chiral symmetry is
restored. Dileptons and photons can also give information on the early
temperature, via the black body radiation spectrum.

\begin{figure}[hbt]
  \begin{center}
    \hspace*{-0.12in}
    \includegraphics[width=1.\linewidth]{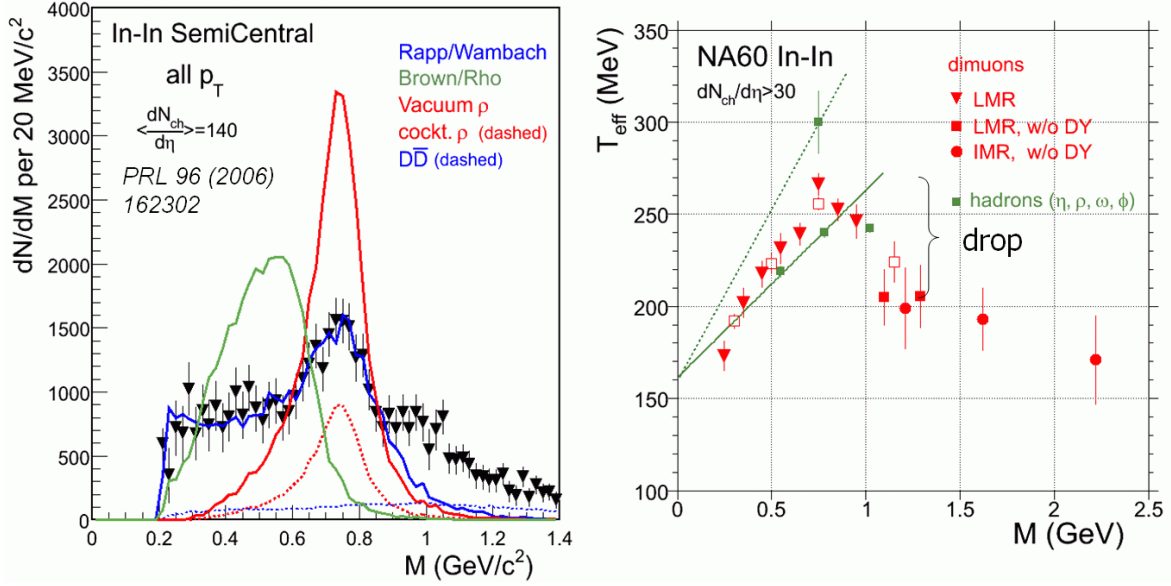}
  \end{center}\vspace*{-0.12in}
  \caption{\label{fig:NA60} Left: The dimuon invariant mass spectrum
    in semi-central In+In collisions measured by NA60, as compared
    to models. Right: Inverse slope parameter for various mass bins.
  }
\end{figure}

Leaving aside the question of whether the sQGP is formed at the SPS or
not, we turn to the NA60 experiment which has made a precision
measurement of the low mass dimuon spectrum in semi-central In+In 
collisions\cite{Arnaldi:2006jq}(see Fig.~\ref{fig:NA60}, left).  Note
that the atomic weight of In is 115, about half the atomic weight of
lead or gold used in previous SPS experiments and RHIC.  The model of
Rapp and Wambach where the $\rho$ spectral function is modified fits
the invariant mass spectrum - the data is consistent with a broadening
of the $\rho$ with no mass shift.  This is an important result, as
hadronic models are essentially an attempt to do many body QCD
calculations below T$_C$ and include an approach to the restoration of
chiral symmetry as T$_C$ is approached from below (see
\cite{Rapp:2006rh}). Theoretical calculations show that baryons are
important for this effect. When baryons are taken out of the
calculation, the models are no longer able to explain the data. It
then is reasonable to conjecture that the effect seen in NA60 is a
result of the baryon density at the SPS. Generally the modification of
particle masses and widths are a slow function of the baryon density
below the critical temperature. Even at the baryon density of nuclear
matter, meson masses and widths are expected to be modified\footnote{In contrast, as a function of temperature, the modification of hadrons is
rather sharp in the region of T$_C$ for low baryon density. 
It is interesting to note, however, that in many of
these models, even in RHIC collisions where the net baryon density is low, 
broadening of the meson masses
is due to baryons - however in this case it is due to the non-equilibrium
presence of baryon and their corresponding antibaryons\cite{Rapp:2002mm}.}.

In the model, the excess on the high mass side of the $\rho$ is not
explained.  In order to study this region, the NA60 collaboration
divided the data into sub-samples of varying invariant mass, and then
fitted the p$_T$ spectra of the sub-sample. The resultant slopes are
shown in Fig.~\ref{fig:NA60} (right). They note that for masses below
about 800 MeV - i.e. below the $\rho$ - the slope of the p$_T$ spectra
rises. This is reminiscent of the mass dependent p$_T$ spectra of
hadrons - an indicator of radial flow of a hadronic source i.e.
primarily the $\rho$. The radial flow is built up in the later stages
of the collision which is dominated by hadrons. Above $\sim$800 MeV,
the slopes drop, which could be an indicator that the source of at least some of  such
dileptons are from earlier in the collision, i.e. before the radial
flow develops\cite{Arnaldi:2007ru}. This might imply that the source is of
a partonic nature.

One of the new results shown at this conference by NA60 was an analysis of
the polarization by examining the polar angle of the excess dimuons 
in the Collins Soper Frame. Within errors the distribution is consistent
with an unpolarized source as expected of thermal radiation\cite{Damjanovic:2008}. 

\begin{figure}[hbt]
  \begin{center}
    \hspace*{-0.12in}
    \includegraphics[width=0.5\linewidth]{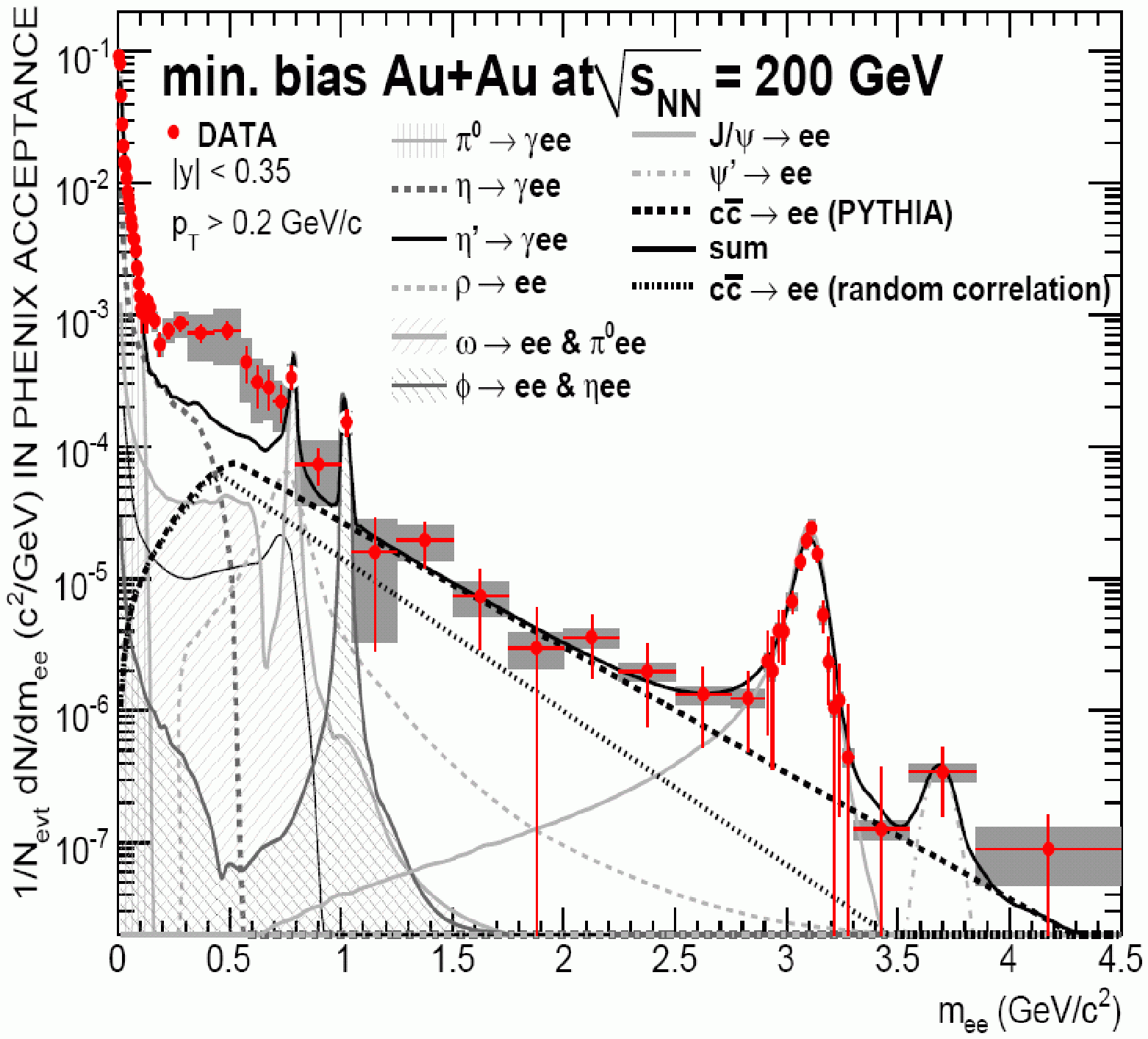}
    \includegraphics[width=0.5\linewidth]{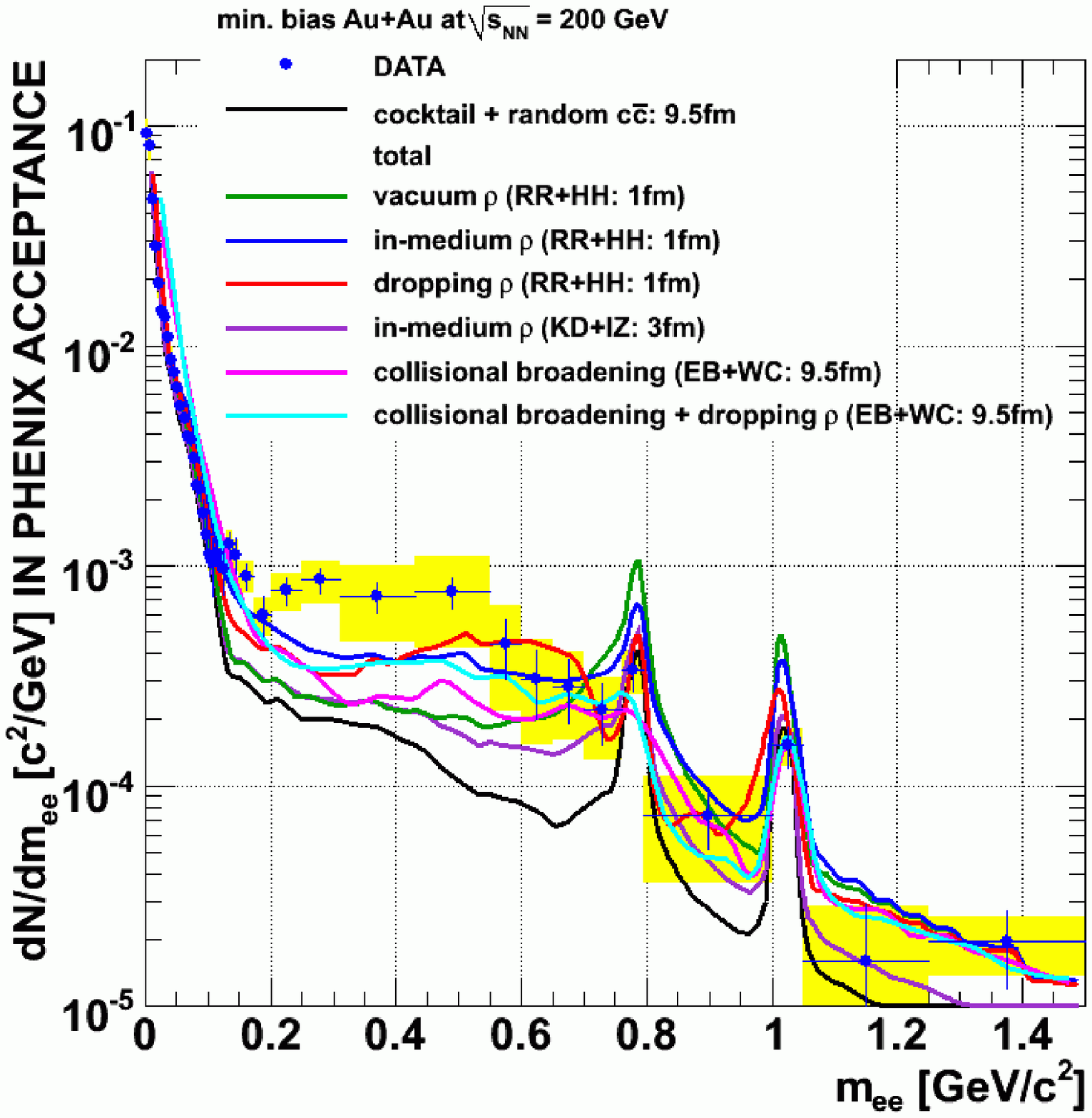}
  \end{center}\vspace*{-0.12in}
  \caption{\label{fig:mee} Left: The dielectron mass spectrum as
    compared to p+p and hadronic cocktail normalized at
    m$_{ee}<20$~MeV/c$^2$ as measured by PHENIX in p+p and Au+Au at
    200 GeV. Right: The dielectron invariant mass spectrum compared
    to models.  }
\end{figure}

\begin{figure}[hbt]
  \begin{center}
    \hspace*{-0.12in}
    \includegraphics[width=1.\linewidth]{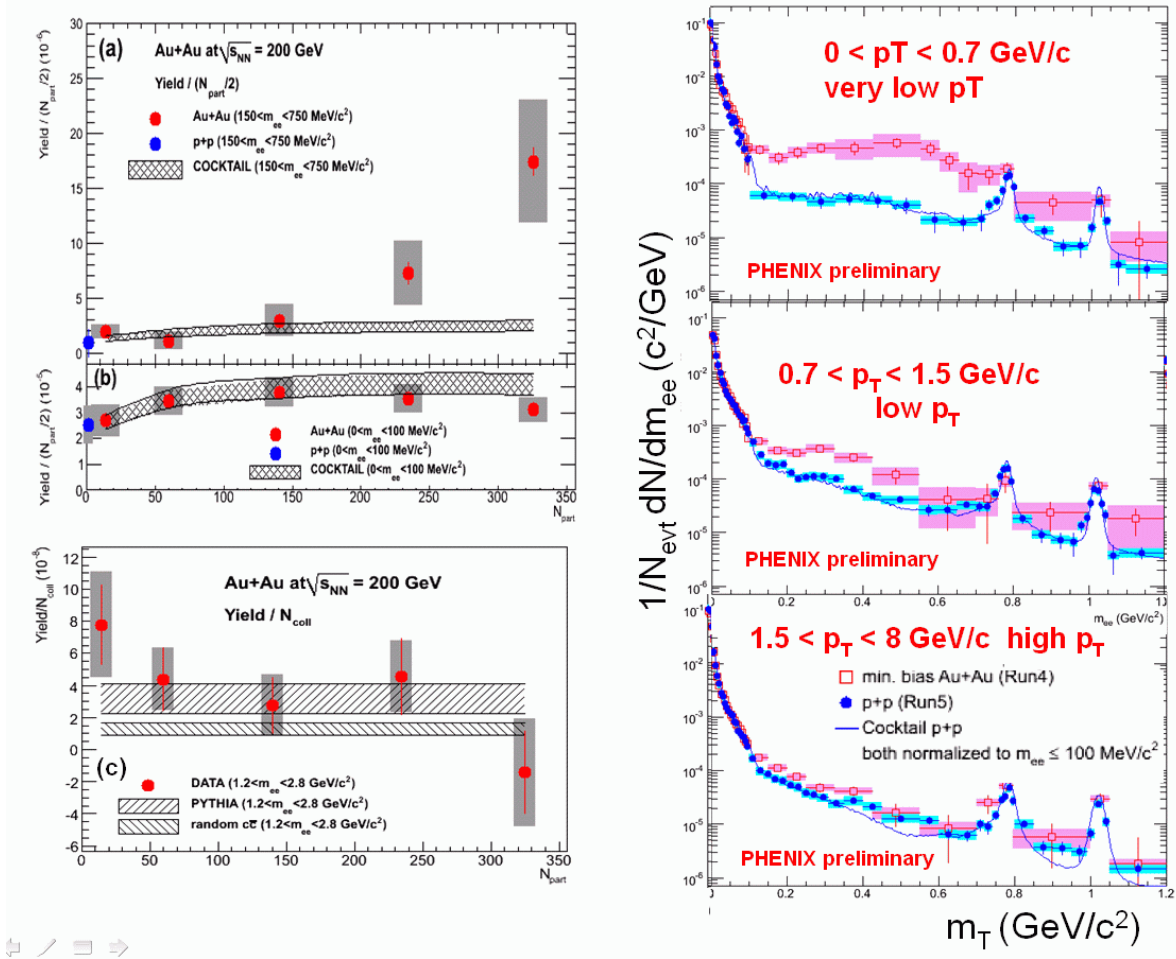}
  \end{center}\vspace*{-0.12in}
  \caption{\label{fig:mee_cent_pt} Left: Centrality dependence of
    dielectron excess in various mass regions as measured by PHENIX
    in 200 GeV Au+Au collisions. The top panel is for the low mass
    region and is normalized to N$_{part}$. The center panel is for
    the $\pi^0$ mass region where the data is normalized to the
    cocktail. The bottom panel shows the high mass region normalized
    to the number of binary collisions. Right: The invariant mass
    spectrum in 3 regions of p$_T$.  }
\end{figure}

Now we turn to the RHIC data. In particular, since we have strong
evidence for thermalization above the critical temperature, and since
there is no doubt that QCD is the governing theory, we must detect
thermal radiation, and chiral symmetry must be restored. What we would
like to know is, a) a measure of the temperatures involved and b) the
mechanism and consequences of the restoration of chiral symmetry. At
this conference we have seen two crucial results from the PHENIX
collaboration on this subject.  The first is the observation of an
excess in the dilepton spectrum over what would be
predicted from a hadronic cocktail as shown in
Fig.~\ref{fig:mee}(left). In the figure, the normalization of the p+p
and cocktail spectra to the Au+Au data is done in the region of the
$\pi^0$. Fig.~\ref{fig:mee}(right) shows that the $\rho$ broadening
model which was successful in explaining the NA60 data no longer works
in the region of invariant mass between 150 and 750 MeV. So far,
no model appears able to explain the data. We then look
at the general behavior of the excess as a function of centrality and
transverse momentum. Fig.~\ref{fig:mee_cent_pt}(left) shows the
centrality dependence of the excess in three mass regions: the $\pi^0$
region where the data is normalized, a low mass region between 0.15
and 0.75 GeV/c$^2$, and a high mass region between 1.2 and 2.8
GeV/c$^2$. In the high mass region, the excess scales with binary
collisions, and agrees with a cocktail which includes charm. In the
low mass region, the excess increases with centrality.
The p$_T$ dependence of the low mass
region is shown in Fig.~\ref{fig:mee_cent_pt}(right). The excess is
strongest in the lowest p$_T$ bin (0-0.7 GeV/c) and is reduced at the
highest p$_T$.  The fact that the excess is strongest in the most
central, low p$_T$ collisions is consistent with the signal coming
from a modified hadron in the fireball. Particles at low p$_T$ from
central events, would have presumably been exposed for the longest
time to the highest energy density region of the collision. The data
is subdivided into bins of invariant mass, and the inverse slopes of
the p$_T$ distribution are fit. Since there is clearly a strong p$_T$
dependence which may indicate different sources, the fit is done in a
low p$_T$ region of $0<m_T<1$~GeV/$c^2$ and an intermediate p$_T$
region of $1<m_T<2$~ GeV/$c^2$.  The high p$_T$ region (with large error bars) 
gives a rising
inverse slope consistent with the source being the radial flow of
modified hadrons (the $\rho$?), albeit as mentioned previously,
models including this effect cannot completely describe the excess and
alternate explanations may be in order.
The low p$_T$ region fits to a slope of about 120 MeV/c$^2$ which seems 
low to be from a partonic source in
the sQGP since it is below any of the standard values of T$_C$ from
the lattice\cite{Toia}.

\begin{figure}[hbt]
  \begin{center}
    \hspace*{-0.12in}
    \includegraphics[width=.6\linewidth]{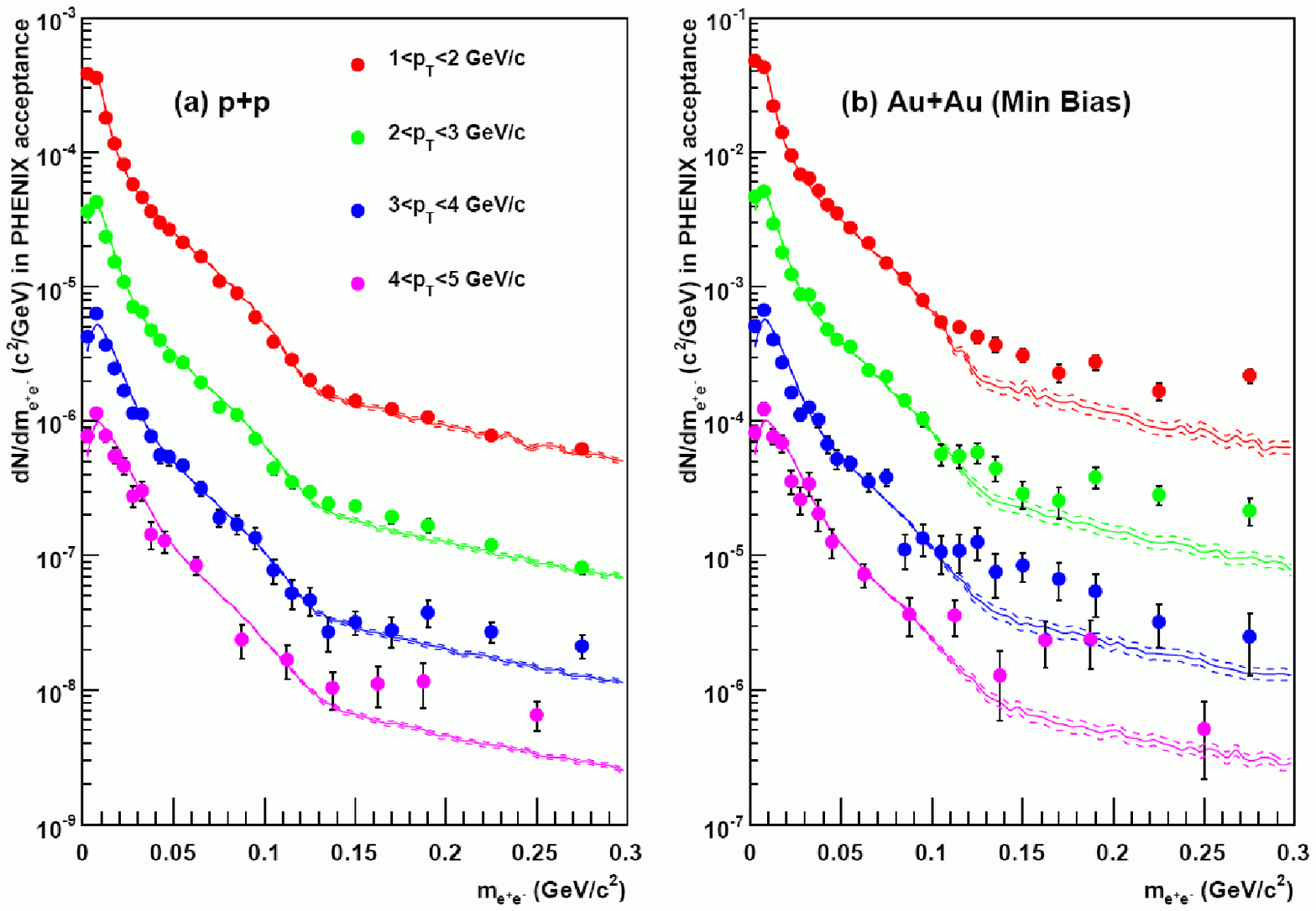}
    \includegraphics[width=.4\linewidth]{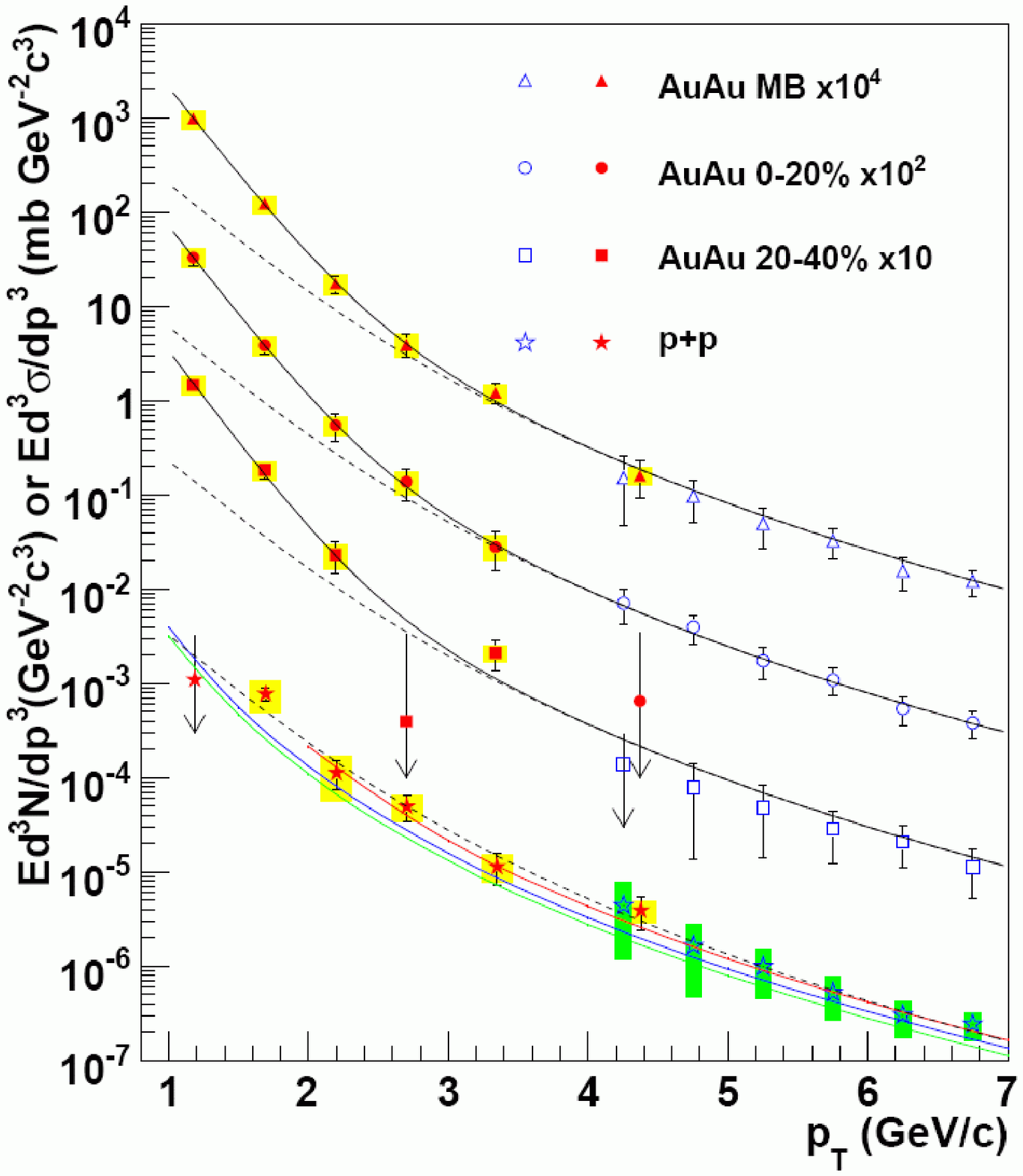}
  \end{center}\vspace*{-0.12in}
  \caption{\label{fig:mee_virt} Left 2 panels: the dielectron
    invariant mass spectrum in 200 GeV p+p and Au+Au collisions as
    compared to a hadronic cocktail. Right panel: Invariant cross
    section (p+p) and invariant yield (Au+Au) of direct photons as a
    function of p$_T$.  The three curves on the p+p data represent NLO
    pQCD calculations, and the dashed curves show a modified power-law
    fit to the p+p data, scaled by T$_{AA}$. The black curves are
    exponential plus the T$_{AA}$ scaled p+p fit. }
\end{figure}

A second new result is the p+p confirmation to the background of the
excess in virtual direct photons as detected through the dilepton
channel shown at QM2005. Photon measurements are very difficult due to
the large background from photons from $\pi^0$s. A way to overcome
this is to look at virtual photons which internally convert into
dielectrons. If one cuts on an invariant mass above the $\pi^0$ mass,
this effectively suppresses the contribution from the $\pi^0$ decay.
Fig.~\ref{fig:mee_virt}(left two panels) shows the dilepton invariant
mass as compared to the cocktail. The cocktail and data are normalized
in the region m$_{ee}<30$~MeV.  For p+p collisions the cocktail and
data match well, aside from a small excess at high p$_T$ (see
Fig.~\ref{fig:mee_virt}, left). In Au+Au collisions one sees an excess
when one looks above the $\pi^0$ mass.  The data is then fit to a
combination of cocktail and a spectrum assumed to come from the
internal conversion of direct photons. The factor r which governs the
strength of the internal conversion component is free to vary. If one
compares the measured value of r to that expected from a NLO pQCD
calculation the p+p data nicely match, while the Au+Au data show a
large excess (see \cite{Dahms:2008}). The fraction of direct photons
is then multiplied by the inclusive photon yield and a cross section
is produced (see Fig.~\ref{fig:mee_virt}, right). A fit to the
spectrum from the 20\% most central collisions to an exponential
plus a T$_{AA}$-scaled p+p fit function, yields an inverse slope of
221$\pm$23$\pm$18 MeV. Assuming that the direct photons are of a
thermal origin, the extracted inverse slope can be related to an
initial temperature T$_{init}$.  Hydrodynamic models which include the
evolution of the system yield between 300 and 600 MeV for the initial
temperature, where T$_{init}$ is strongly dependent on the initial
thermalization time; as one might expect, later times yield lower
temperatures\cite{Adare:2008anew}.

\section{The J/$\psi$}
\begin{figure}[hbt]
  \begin{center}
    \hspace*{-0.12in}
    \includegraphics[width=0.7\linewidth]{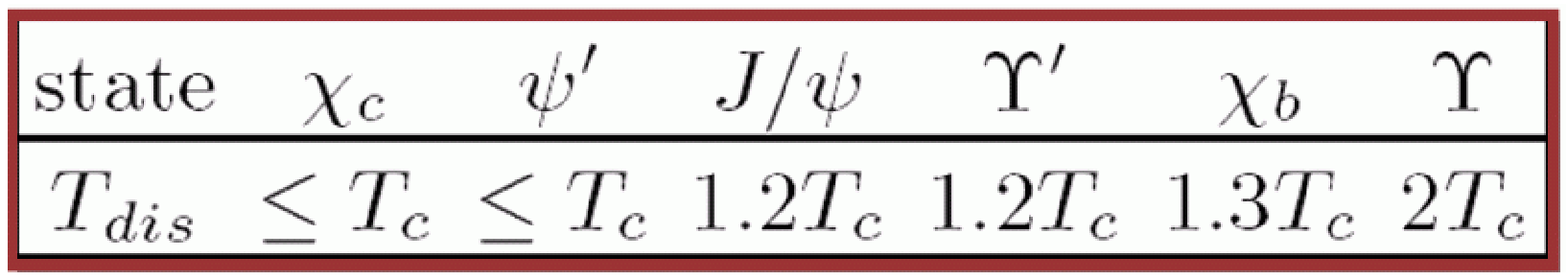}
  \end{center}\vspace*{-0.12in}
  \caption{\label{fig:Tmelt} 
     Quarkonium dissociation temperatures from \cite{Mocsy:2007jz} .
    }
\end{figure}

The J/$\psi$ data at RHIC presents a conundrum. Is J/$\psi$
suppression a sure sign of deconfinement? At RHIC, the possibility of
production through coalescence is an important consideration, as is
the initial state of the nucleus - the Colored Glass Condensate, or
shadowing, absorbed under the general heading of ``cold nuclear
effects''. While this is an experimental review, one of the
encouraging new results shown at this meeting for experimentalists was
theoretical. For the past few years potential models which are used to
make predictions for heavy ion collisions and lattice results have not
agreed. At this conference agreement was shown between a potential
model and lattice correlators for all states\cite{Mocsy:2007yj,Mocsy:2007jz}. This goes a long way to putting the suppression
via deconfinement models on stronger footing. The ``melting
temperatures'' of the various charmonium have changed from previous
values (see Fig.~\ref{fig:Tmelt}).  The possibility of coalescence and
initial state effects are still present. However, it is clear that charmonium
can tell us something about deconfinement, if we can disentangle
deconfinement from other effects. This will require both improvements
to dynamic model calculations including coalescence, and good
experimental data sets in proton-nucleus or deuteron-nucleus collisions
to understand cold nuclear matter effects.

\begin{figure}[hbt]
  \begin{center}
    \includegraphics[width=1.1\linewidth]{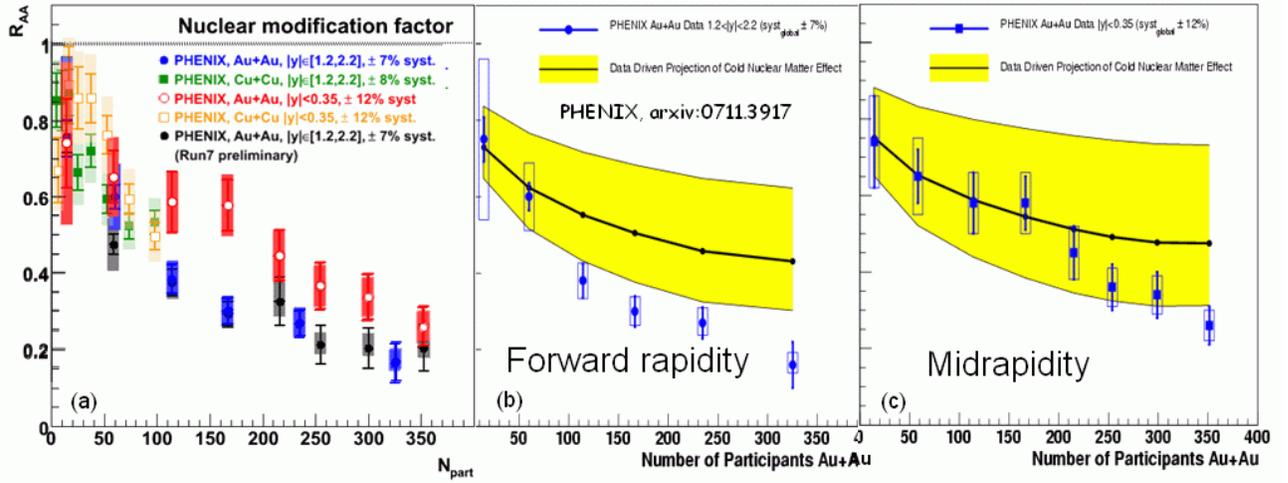}
  \end{center}\vspace*{-0.12in}
  \caption{\label{fig:jpsisupp} Left panel: R$_{AA}$ for the J/$\psi$
    as measured by PHENIX. Note that the forward rapidity region shows
    more suppression than the mid-rapidity region. Right two panels:
    The forward and mid rapidity data together with a projection of
    the cold nuclear matter effect, using PHENIX d+Au data.  }
\end{figure}

Data from PHENIX including Cu-Cu results, and a first look at a larger
sample of Au+Au data are consistent with previous results (see
Fig.~\ref{fig:jpsisupp}, left).  Naively one would think the energy
density is lower in the forward rapidity region, therefore the effect
of deconfinement would be weaker. Just the opposite is seen.
Suppression at forward rapidity is \it{stronger} \rm{} than at
midrapidity!  One then turns to the d+Au data to understand the effect
of the initial state on the suppression. Fig.~\ref{fig:jpsisupp}(right
two panels) shows the projection of the cold nuclear matter effect,
taken from d+Au measurements. A data driven technique is used in which
it is assumed that a single modification factor which is a function of
the radial position in the nucleus parametrizes all cold nuclear
matter effects\cite{Adare:2007gn}. Unfortunately the quality of the
d+Au data is too poor to give a conclusive answer. A recently obtained
data set should give PHENIX a sample 30 times as large which should go
a long way to pinning down the cold nuclear matter effects. Future
measurements of the $\psi'$, $\chi_C$, and $\Upsilon$ states, together
with open charm and bottom measurements in p+p, d+Au, and Au+Au
collisions should provide enough information to tease apart the the
different phenomena - i.e. cold nuclear effects, coalescence, and
deconfinement.

\begin{figure}[hbt]
  \begin{center}
    \hspace*{-0.12in}
    \includegraphics[width=1.\linewidth]{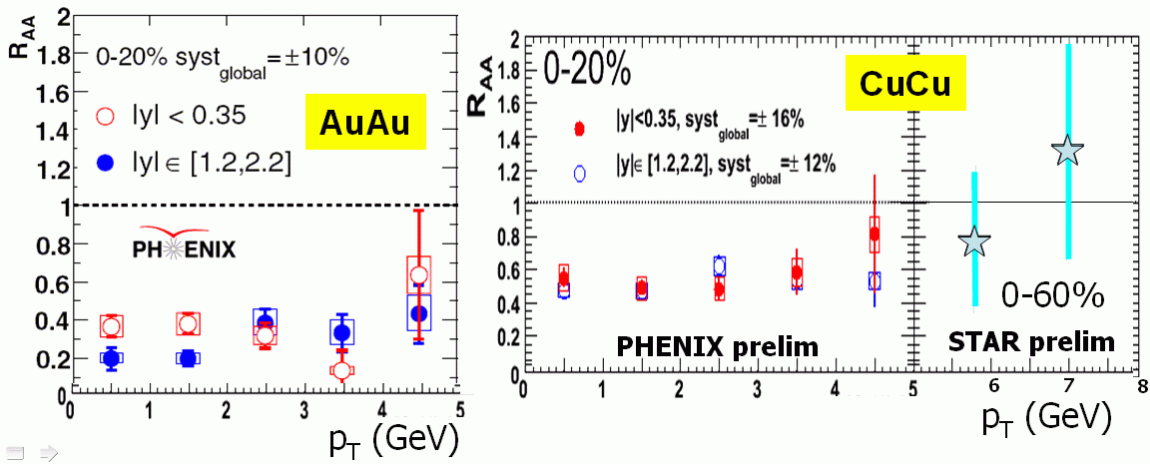}
  \end{center}\vspace*{-0.12in}
  \caption{\label{fig:jpsihipt} The p$_T$ dependence of the J/$\psi$
    R$_{AA}$ as measured by PHENIX and STAR.  }
\end{figure}

One of the interesting predictions coming into this conference from
AdS/CFT is that the suppression of the J/$\psi$ would increase with
p$_T$\cite{Liu:2006nn}. A two component model which includes
dissociation and regeneration\cite{Zhao:2007hh} predicts a flat p$_T$
with coalescence dominating at low p$_T$ and dissociation at high
p$_T$. Early predictions based on a simple potential model\cite{Blaizot:1987ha} predict less suppression as p$_T$ is
increased. STAR has made a first attempt to look at the very high
p$_T$ dependence of the J/$\psi$ via its decay to dielectrons
identified in a combination of the TPC and EMC and triggered by the
EMC (Fig.~\ref{fig:jpsihipt}). The figure also shows the PHENIX
data. At the moment the results are inconclusive, but it will be
interesting to see the dependence of the J/$\psi$ on transverse
momentum once the two collaborations get higher statistics data. This
depends critically on the available running time at RHIC to obtain
good statistics at high p$_T$.

\section{Conclusion}
In closing there were two main results in the realm of leptons,
photons and heavy quarks that we should take away from this
conference:
\begin{itemize}
\item Bottom quarks flow and are suppressed at high p$_T$ in heavy ion collisions.
\item We are now beginning to see a signal for enhanced photon and dilepton
production in heavy ion collisions.
\end{itemize}

\noindent There are clearly questions to be answered in the coming years:
\begin{itemize}
\item What are the sources of the excess dileptons/photons?
\item Are they telling us the temperature? What is it?
\item How do the heavy quarks manage to become part of the bulk? How
strong is the ``s'' in sQGP. What is the viscosity?
\item What is the mechanism of the rapid thermalization, and the
thermalization or partial thermalization of heavy quarks?
\item Does the J/$\psi$ show anomalous suppression due to deconfinement?
\end{itemize}

It is an exciting time for our field. We now have many powerful tools
at our disposal- e.g. the flow of the charmonium to test models of
coalescence and deconfinement, correlations measurements between
photons and jets, or charm quarks and jets, precision measurements of
the dilepton and photon spectra. In addition the LHC will be coming
on in the next few years bringing a host of new information from
higher energies. I would like to close with one important fact. All
of the heavy ion experiments both at RHIC and at the LHC depend and
will depend on adequate running time, and support for the detectors
and analysis. The possibilities ahead of us are tremendous, but we
must ensure that support for the field does not waver, else we will
leave many of these questions unanswered.

\end{document}